\documentclass[aps,reprint,10pt,twocolumn,superscriptaddress]{revtex4}
\usepackage{amsmath}
\usepackage{amssymb}
\usepackage{bbm}
\usepackage{graphicx}
\usepackage[countmax]{subfloat}
\usepackage{framed}
\usepackage{color}
\usepackage{hyperref}
\usepackage{epstopdf}
\usepackage{dsfont}
\usepackage{amsthm}
\usepackage{wrapfig}
\usepackage{relsize}
\usepackage{bm}
\usepackage{enumitem}
\usepackage{soul}

\newcommand{\red}[1]{{\color{red}}}

\usepackage{pifont}

\newcommand{\ba}{\begin{eqnarray}}
\newcommand{\ea}{\end{eqnarray}}

\begin{document}

\title{Calibration of quantum sensors by neural networks} 

\author{Valeria Cimini}
\affiliation{Dipartimento di Scienze, Universit\`{a} degli Studi Roma Tre, Via della Vasca Navale 84, 00146, Rome, Italy}
\author{Ilaria Gianani}
\affiliation{Dipartimento di Scienze, Universit\`{a} degli Studi Roma Tre, Via della Vasca Navale 84, 00146, Rome, Italy}
\affiliation{Dipartimento di Fisica, Sapienza Universit\`{a} di Roma, P.le Aldo Moro, 5, 00185, Rome, Italy}
\author{Nicol\`o Spagnolo}
\affiliation{Dipartimento di Fisica, Sapienza Universit\`{a} di Roma, P.le Aldo Moro, 5, 00185, Rome, Italy}
\author{Fabio Leccese}
\affiliation{Dipartimento di Scienze, Universit\`{a} degli Studi Roma Tre, Via della Vasca Navale 84, 00146, Rome, Italy}
\author{Fabio Sciarrino}
\affiliation{Dipartimento di Fisica, Sapienza Universit\`{a} di Roma, P.le Aldo Moro, 5, 00185, Rome, Italy}
\author{Marco Barbieri}
\affiliation{Dipartimento di Scienze, Universit\`{a} degli Studi Roma Tre, Via della Vasca Navale 84, 00146, Rome, Italy}
\affiliation{Istituto Nazionale di Ottica - CNR, Largo Enrico Fermi 6, 50125, Florence, Italy}

\begin{abstract} 
Introducing quantum sensors as solution to real-world problem demands reliability and controllability outside laboratory conditions. Producers and operators ought to be assumed to have limited resources ready available for calibration, and yet, they should be able to trust the devices. Neural networks are almost ubiquitous for similar tasks for classical sensors: here we show the applications of this technique to calibrating a quantum photonic sensor. This is based on a set of training data, collected only relying on the available probe states, hence reducing overheads. We found that covering finely the parameter space is key to achieve uncertainties close to their ultimate level. This technique has potential to become the standard approach to calibrate quantum sensors.
\end{abstract}

\maketitle

Quantum technologies are experiencing a world-wide effort to foster their applications beyond what is achieved in a laboratory. In particular for quantum sensing, quantum resources are promising to reach accuracy beyond what is permitted from classical counterparts~\cite{PhysRevLett.96.010401}. This advantage however, it is conditioned on a robust operation in the presence of noise as well as imperfections of the measuring instruments~\cite{Dooley:2018vl,PhysRevA.94.042101}. Many methods have been proposed to this end, including error correction~\cite{PhysRevLett.116.230502,Zhou:2018jt,Reiter:2017sf,PhysRevA.94.012324}, monitoring of the environment~\cite{Albarelli}, and imperfection-tollerant probe state design~\cite{Kacprowicz:2010,PhysRevLett.107.113603}. 

Regardless the method adopted, it is vital to devise analysis methods that grant optimal use of the collected data for estimation, {i.e.} to obtain the so called optimal estimator. For simple instances, the maximum likelihood approach or methods based on Bayes' theorem are known to provide such an estimator~\cite{PhysRevLett.109.180402,PhysRevApplied.10.044033}. On the other hand, these are generally computationally intensive, often require a more involved characterization of the system~\cite{Lundeen:2008fk,Zhang:2012,PhysRevA.92.032114,Roccia:18}, and thus pose difficulties in scaling to configurations with increased complexity. Further, characterization is generally based on preparing quantum states with different requirements than those actually used in the estimation routine~\cite{PhysRevLett.108.253601,PhysRevX.4.041025,PhysRevLett.116.100802}. In the perspective of compact architectures, the resulting requirement of flexibility may come at odds with that of reliability and reduced costs~\cite{PhysRevLett.118.190502}. A method being self-consistent, resource economic, and versatile is desirable. 

Nowadays, the incredible amount of data collected in diverse problems requires efficient self-adjustment of the sorting protocol. The size and complexity of these problems has imposed Machine Learning (ML) algorithms as the mainstream solution in these situations \cite{Murphy2012,Simon2013}. 
ML has been recently proposed and applied as tool for characterization and optimization of quantum systems as well as for handling quantum physics problems \cite{Dunjko18}. Notable examples include its adoption for the learnability of quantum measurements, states and processes \cite{PhysRevLett.114.200501,Aaronson2007, Rocchettoeaau1946,Yu2019,Spagnolo17,Granade_2012,Wang17}, validation of multiparticle interference \cite{Flamini19,PhysRevX.9.011013}, quantum state engineering \cite{PhysRevA.96.062326,PhysRevLett.122.020503, PhysRevX.8.031086}, and as a tool for quantum experiment design and control \cite{Briegel12,Innocenti2018,PhysRevLett.116.090405,Melnikov1221,PhysRevLett.116.230504,PhysRevA.95.012335}.
In the context of quantum metrology, ML has found an application in quantum phase estimation protocols to efficiently extract the information encoded in the probe \cite{Wiebe2015b, Wiebe2016a, Paesani2017}. More specifically, ML represents a powerful toolbox to optimize, via adaptive protocols \cite{Hentschel1,Hentschel2,Lovett13} the performance of a sensor operating with a small collection of repetitions~\cite{PhysRevApplied.10.044033}.
These considerations suggest the viability of this approach for a full characterization of quantum sensing apparata. Specifically, neural networks can extract an output value of the parameter of interest, following their training on a set of inputs associated to a calibrated set of parameters; this is an efficient algorithm that can be run on ordinary machines \cite{Leccese1, Leccese2}. No explicit modelling of the imperfections is thus needed, as that information will be taken into account, although in implicit form, in the training itself.  On the other hand, the training data can only be collected in finite time, hence with finest statistics: it is important to understand how the associated uncertainty influences the quality of the final estimation, and its capability of showing  quantum enhancement. 

\begin{figure}[b]
\includegraphics[width=\columnwidth]{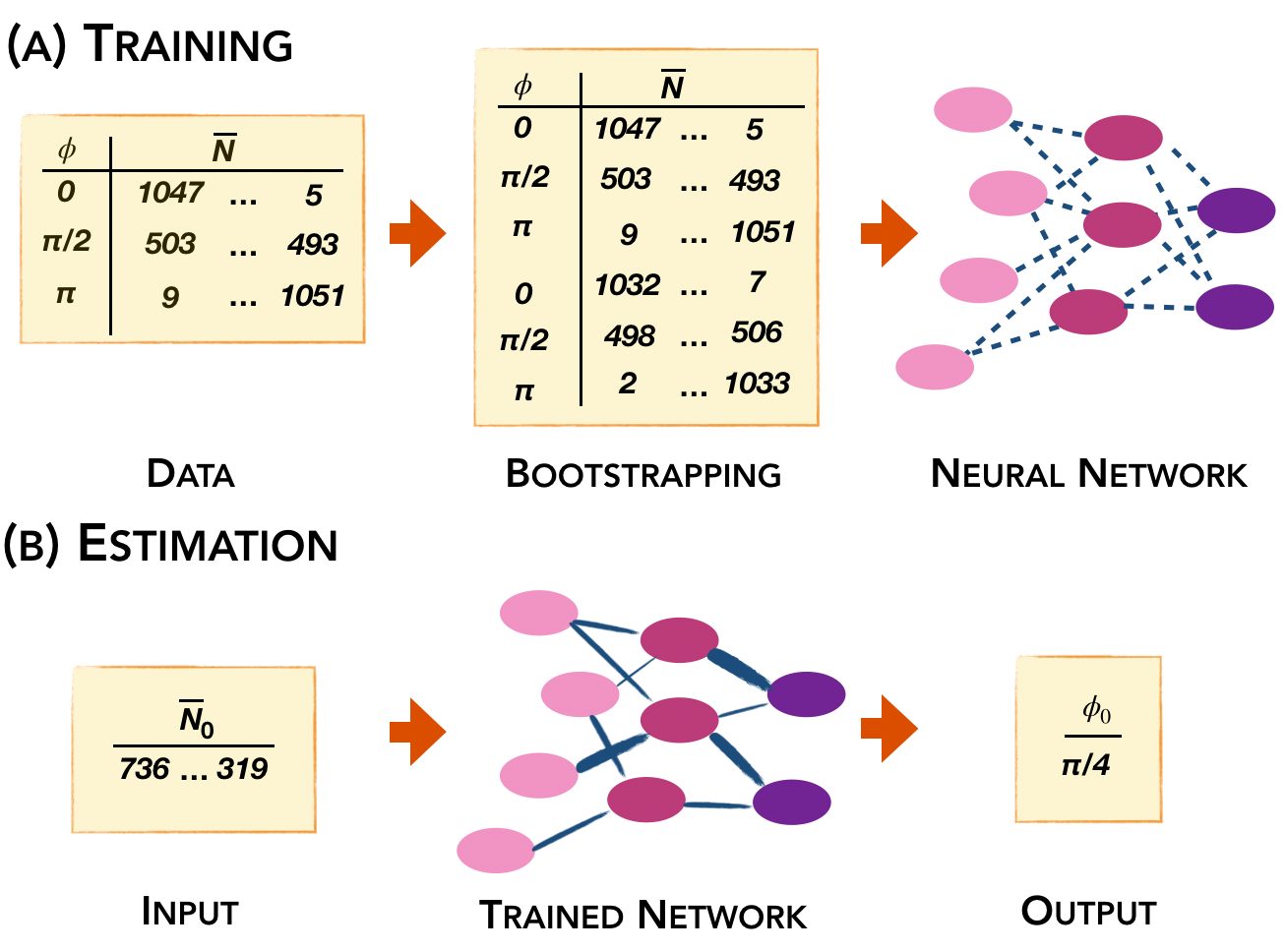}
\caption{Schematics of the use of neural networks for parameter estimation. The first step (a) consists in training the network by inputting a set of test data using bootstrapping to account for uncertainties. Upon completion, the actual estimation (b) uses the trained network to extract an estimate of the parameter.}
\label{fig:figure1}
\end{figure}

In this article, we discuss the characterization of a quantum phase sensor based on N00N states by means of the neural network. We find that the simplicity of the characterisation does not impact heavily the metrological capabilities of the device. The algorithm has no strict requirements in its settings to provide near-optimal performance of the estimation, provided that the set of the parameter is explored with adequate resolution. Thanks to its compact implementation and its scalability, this method can find wide applications in future quantum technologies.

The concept behind our investigation is depicted in Fig.~\ref{fig:figure1}. The training step (Fig.~\ref{fig:figure1}a) consists in collecting a set of data $\vec{N}(\phi)$ corresponding to different values of the parameter of interest $\phi$; in general $\vec{N}$ will be in the form of a vector, since it contains multiple measurements, as needed to obtain a correct normalization, to remove ambiguities, or to account for multiple parameters. This is used as an input to a network, consisting of a set of neurons connected among them, possibly forming subsequent layers. The training procedure establishes the weights associated to the connections between each pair of neurons. Unavoidably, an uncertainty will be associated to these measured data, and the network needs to be trained to account for this variability. If the noise statistics on each measurement in $\vec{N}$ is known, a bootstrapping method can be employed to generate multiple fictitious runs of the experiment by means of a Monte Carlo routine. For a fixed network size, the quality of the training will be influenced by the resolution at which $\phi$ is sampled, as well as the number of repetitions of $M$ used for each value of $\phi$. Once the training is complete, the device can be used for parameter estimation: the network is now operated to accept the collected data $\vec{N}_0$ as an input, and to provide an estimation $\phi_0$ as the output (Fig.~\ref{fig:figure1}b). By using the same bootstrapping method above on $\vec{N}_0$, the uncertainty $\Delta\phi_0$ can also be evaluated. 

\begin{figure}[t]
\includegraphics[width=\columnwidth]{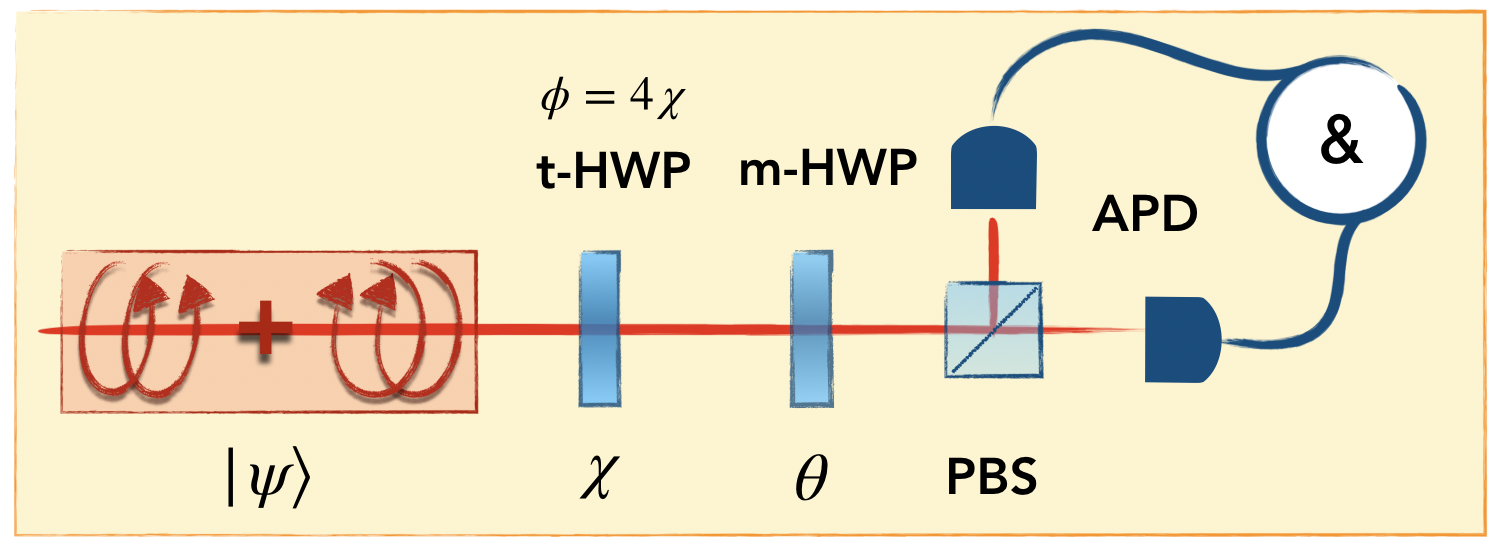}
\includegraphics[width=\columnwidth]{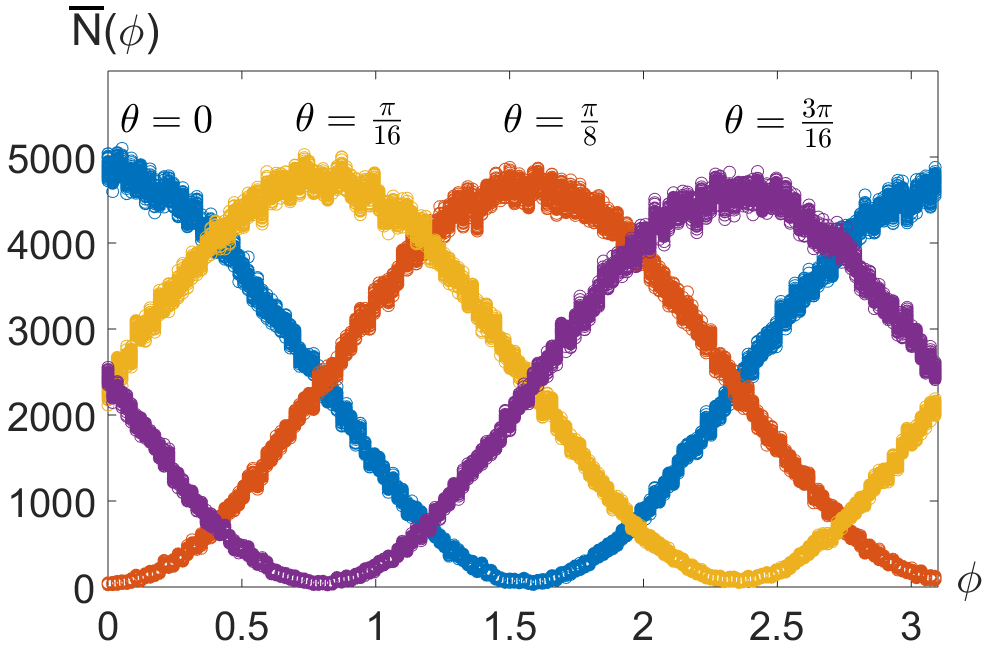}
\caption{Top: experimental setup. The N00N state probe accumulates a phase $\phi=4\chi$ by means of a test-half waveplate (t-HWP). The analysis is carried out by a second measurement plate (m-HWP), a polarising beamsplitter, and coincidence detection of two avalanche photodiodes (APDs). Bottom: Part of the experimental training set. The registered count rates and those derived from bootstrapping are reported as a function of the calibrated phases $\phi$. The labels indicate the angular setting $\theta$ of the measurement HWP, corresponding to four different polarisation projections.}
\label{fig:figure2}
\end{figure}

We test this method in a quantum phase estimation experiment. A two-photon N00N state on the $R$ight- and $L$eft-circular polarisation of a single spatial mode approximating $\vert \psi \rangle{=}2^{-1}[(a^\dagger_R)^2+(a^\dagger_L)^2]\vert0\rangle$ is used for this task (see Fig. \ref{fig:figure2}). This is achieved by overlapping on the same spatial mode two otherwise-indistinguishable orthogonally polarized photons by means of a polarizing beam splitter (PBS) \cite{Pryde17,Roccia:18}: the photon pairs are generated via a type-I Spontaneous Parametric Down Conversion (SPDC) source using a 3mm BBO crystal pumped with a 405 nm CW laser; their polarization is then set orthogonal by means of Half Wave plates (HWPs) and they are sent on a PBS obtaining the circular polarization N00N state. The measurement is then carried out collecting a vector of coincidences counts associated to photon pairs after they have accumulated a relative phase $\phi$, resulting from the rotation of their polarization state. For each phase, we look at the coincidences counts relative to four different polarizations projections~\cite{Roccia:18}. 

Even in this simple instance, many imperfections affect the actual experimental apparatus including those linked to the phase preparation as well as imperfect splittings on the PBSs and polarization-dependent efficiencies of the detection channels. Scrupulous modelling should include several variables to account for these in a parameter-dependent way, making the problem involved. In more complex scenarios this becomes even more demanding. We test how the effectiveness of the implicit treatment of imperfections made possible by neural networks.

We collect data for $180$ different phase values from $0$ to $180^\circ$ in steps of $1^\circ$, simply obtained by rotating a half wave plate (HWP) from $0$ to $45^\circ$ in steps of $0.25^\circ$ -- the relation between the phase $\phi$ and the angle setting $\chi$ is $\phi = 4 \chi$. The initial training data thus consist of 180 vectors, each containing four counting rate. Each projection accumulated data for 1s, which, at the observed count rate, correspond to around 10000 total events for each phase, divided among the four projections (Fig.~\ref{fig:figure2}). 

\begin{figure}[b]
\includegraphics[width=\columnwidth]{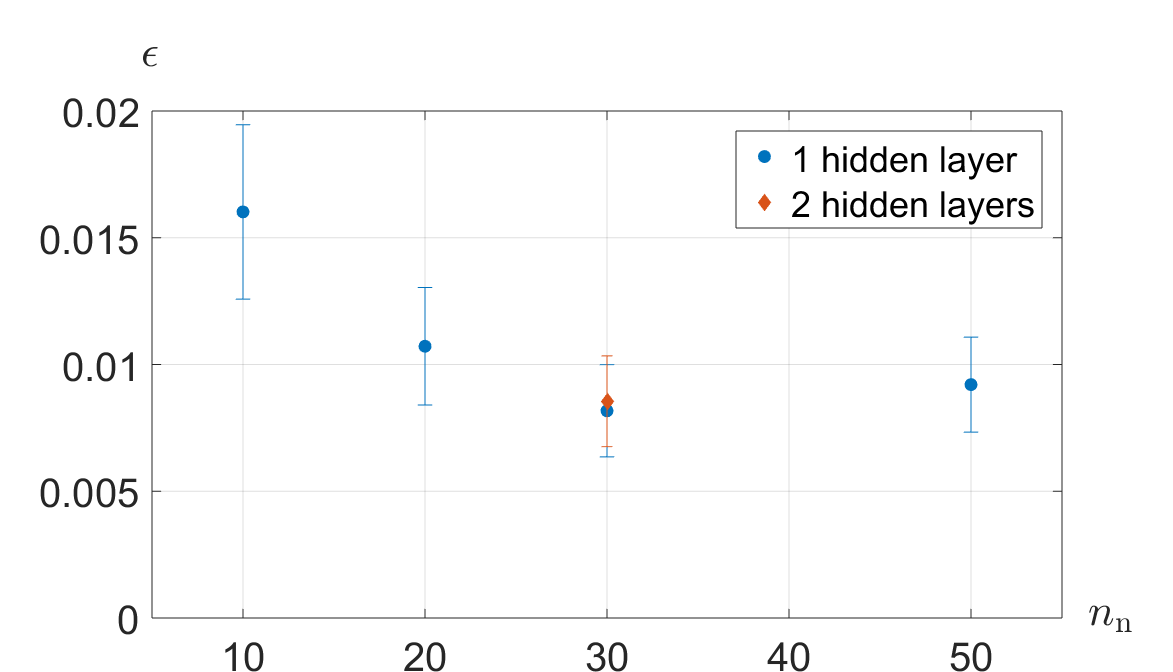}
\includegraphics[width=\columnwidth]{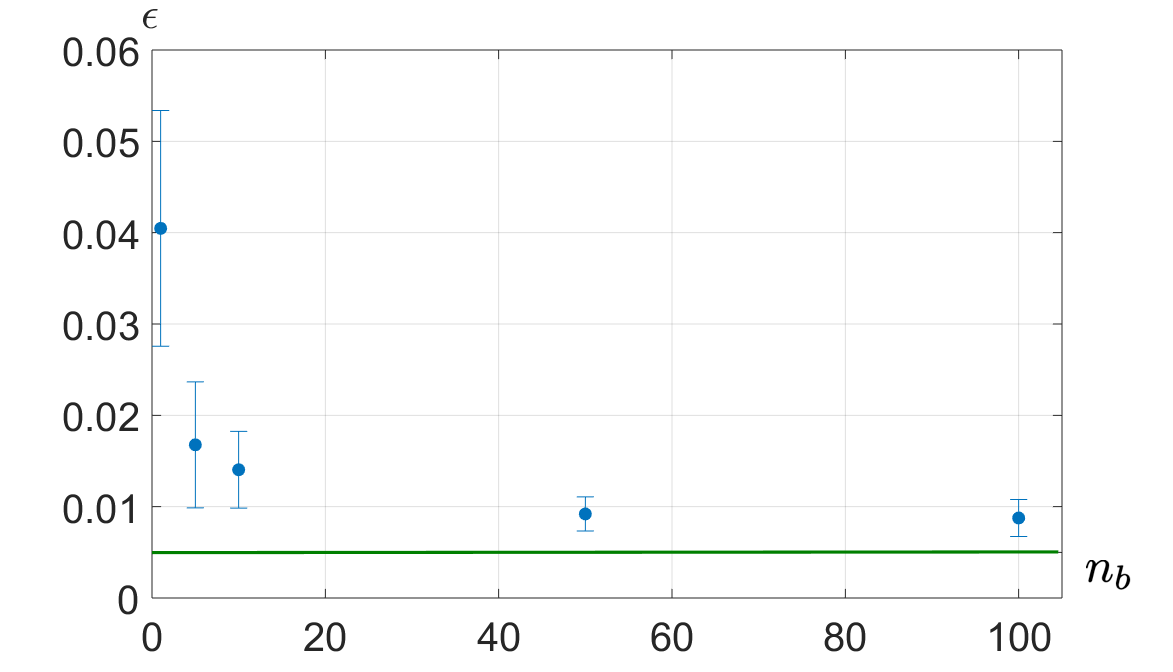}
\caption{Exploration of the training parameters, based on $M=10000$ simulated events. Upper panel: estimation error $\epsilon$ as a function of the number of neurons $n_{\rm n}$ in the hidden layers of the network at $n_{\rm b}=50$. For $n_{\rm n}=30$, we have considered a single layer, or two layers with 20 and 10 neurons. Lower panel: estimation error $\epsilon$ for different number of Monte Carlo repetitions $n_{\rm b}$ obtained from bootstrapping at $n_{\rm n}=30$. The solid line indicates the Cram\'er-Rao bound. In both panels, the error bars are calculated from 35 different trainings; this takes into account the variations due to the random initialisation of the neural network algorithm at the beginning of the procedure.}
\label{fig:training}
\end{figure}

From all the coincidences counts we obtain the relative frequencies associated to that phase, and we use them for a supervised training of the network. In this way, we can use the network independently on the total number of collected events. Since these counts follow a Poissonian distribution, we can test the accuracy by using the bootstrap procedure described above to feed our network with different repetitions associated to each phase. The network is structured as a feed-forward network with sigmoid hidden neurons~\cite{matlab}. This includes an input layer, and output layer, and a hidden layer with $n_{\rm n}$ neurons. The input data are randomly divided in a training set (70\%), a validation set (15\%) and a test set (15\%). The network is trained with Levenberg-Marquardt backpropagation algorithm designed to optimise for every phase, the precision on the validation set using the mean square error metric. The training stops when the validation error stops decreasing.

\begin{figure}[h]
\includegraphics[width=\columnwidth]{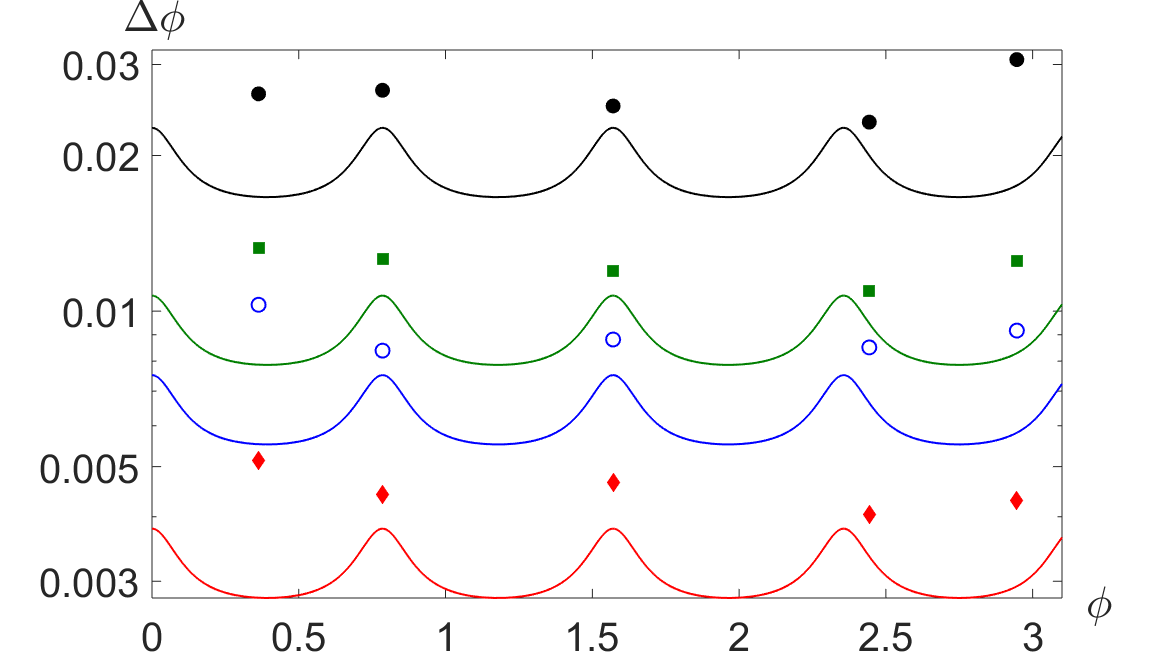}
\includegraphics[width=\columnwidth]{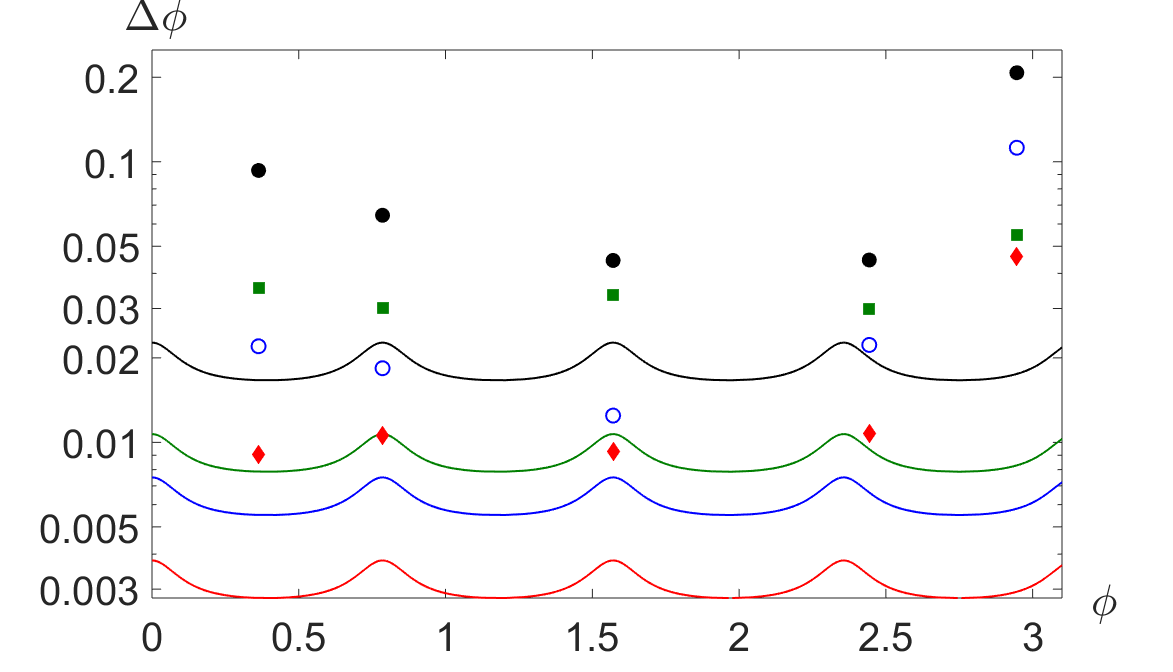}
\caption{Uncertainties for parameter estimation. Upper panel: training set with a sampling step of $1^\circ$. Lower panel: training set with a sampling step of $2^\circ$; notice the logarithmic scale. In both panels, the points refer to total numbers of events $M\simeq1000$ (black solid dots), $M\simeq5000$ (green squares), $M\simeq10000$ (blue empty dots), and $M\simeq40000$ (red diamonds); the curves are the Cram\'er-Rao bounds for the same number of events.}
\label{fig:cramerrao}
\end{figure}

We have run tests on simulated data to establish the optimal parameters for the training for fixed sampling step and total count rate: for this purpose, we input in the trained network the counting rates corresponding to 30 values of the phase, and consider as the error $\epsilon$ the mean standard deviation calculated on 100 repetitions and averaged on all phases. Fig.~\ref{fig:training} shows how this error depends on the number of neurons $n_{\rm n}$ in the network, and on the number of Monte Carlo runs per phase value $n_{\rm b}$. The data do not show sharp optimal working points: we have adopted one hidden layer in the network with 30 neurons, trained with 50 Monte Carlo repetitions.

With the network settings so fixed, we have proceeded to feeding the actual data collected in the experiment for training. We have then collected 5 further sets for the phases 20.8$^\circ$, 45$^\circ$, 90$^\circ$, 140$^\circ$, and 168.8$^\circ$ at different accumulation times (0.5s, 1s, 4s) at the same generation rate as for the training, and 0.5s at 30\% of the initial generation rate. These are used to observe how the uncertainty scales with the number of collected events $M$. In Fig.~\ref{fig:cramerrao}, we show such uncertainties, compared to the associated Cram\'er-Rao bound (CRB)~\cite{PhysRevLett.96.010401,PARIS:2009qy}. The uncertainties remain close to the ultimate limit which also takes into account the reduced contrast of the coincidence oscillations. Remarkably, we observe an improvement also for number of events exceeding those used for the training. The employ of the neural network makes this characterisation robust to noise. We have repeated the training stage decreasing the step size of the phase to $2^\circ$. The distance to the CRB is more pronounced: the network has received insufficient training to produce an output with an accuracy close to optimal. We noticed that, also in this case, a reduction of the uncertainty is observed when increasing the number of events $M$. 
We put these observations in quantitative terms by inspecting $F_{M}=\Delta^2\phi/\sigma^2$, i.e the ratio between the measured variance and the one at the CRB at a given $M$. We obtain the following values: $F_{1000}=1.25$, $F_{5000}=1.39$, $F_{10000}=1.48$, $F_{40000}=1.53$. We note that the F value increases with the repetition number, which implies that variance is diminishing with M however not as fast as the CRB is. This is due to the lack of resolution in the training, which is not a fundamental limit, and in our case it is solely constrained by the accuracy of the actuator of m-HWP.

While overall successful, this characterization underlines some issues which need to be addressed. Close to $\phi{=}0$ and $\phi{=}180^\circ$, a boundary effect can be observed that prevents from obtaining a reliable estimate close to this value; this is removed by increasing the region explored for the training, which should span a wider interval than the one potentially covered in the measurement. Excess uncertainty is associated to phases close to $\phi=90^\circ$, a problem we can attribute to the particular symmetries of the count signals (see Fig.~\ref{fig:figure2}).
We further note that using these four count signals allows to suppress the periodicity ambiguity on the phase in the range 0 - $\pi$, but still leaves an ambiguity in the range 0 - $2\pi$. This latter ambiguity can be approached by integrating adaptive techniques \cite{PhysRevApplied.10.044033} within the phase estimation protocol.

In conclusion, we have applied a neural network algorithm to the calibration of a quantum phase sensor. This method compares favourably to previous investigations that require a complete reconstruction of the functioning of the device~\cite{Lundeen:2008fk,Zhang:2012} or to the data fitting pattern technique~\cite{PhysRevLett.116.100802}. The advantage stems on the one hand from the more efficient data processing, and from the resilience to noise that overcomes the need for regularisation of the data. On the other hand, the characterisation can be carried out only by means of the same kind of states employed in the actual measurement. This is particularly relevant in the perspective of large-scale fabrication of such devices, for which an analysis based on off-line characterisation states would be impractical. 

This advantage is kept also with respect to cases in which an effective characterisation can be carried out in terms of extra parameters -- as long these remain fixed. Indeed, this would rely on a modelling of the imperfections, that might only be captured in part.

Further perspectives of this work can be found in extending the application of neural networks for the calibration of quantum sensors operating in the multiparameter regime, where multiple phases \cite{PhysRevLett.111.070403,Ciampini16,PhysRevLett.119.130504,Polino:19} and relevant system physical quantities~\cite{Roccia:18,Cimini19}, including noise, have to measured simultaneously. Indeed, in those scenario reliable calibration methods becomes particularly crucial due to the increasing complexity of characterising all the system parameters, as well as the computational overhead in handling large amount of data.

{\it Note added. -} During the completion of this manuscript, we became aware of two works~\cite{Macarone2019,Torlai2019}, where neural networks have been applied to quantum state estimation and quantum simulation.

\begin{acknowledgments}
We  acknowledge support  from  the  Amaldi  Research  Center  funded  by the MIUR program  ''Dipartimento di Eccellenza'' (CUP:B81I18001170001).
\end{acknowledgments}

\bibliography{Bibliografia.bib}
\bibliographystyle{apsrev4-1}

\end{document}